\documentclass[showpacs,aps,twocolumn]{revtex4}

\usepackage{bm}
\usepackage{amsmath}
\usepackage{graphicx}
\usepackage{subfigure}
\usepackage[usenames,dvipsnames]{color}
\definecolor{darkblue}{RGB}{0,0,196}
\usepackage[colorlinks=true,linkcolor=darkblue,citecolor=darkblue,urlcolor=darkblue]{hyperref}

\usepackage{setspace}
\usepackage{footmisc}
\usepackage[makeroom]{cancel}
\usepackage{comment}

\def\be{\begin{equation}}
\def\ee{\end{equation}}
\def\ba{\begin{eqnarray}}
\def\ea{\end{eqnarray}}

\usepackage{graphicx}
\usepackage{amsmath,bbm}
\usepackage{amssymb,bm}

\begin{document}

\title{Transverse Momentum Spectra and Nuclear Modification Factor using Boltzmann Transport Equation with Flow in Pb+Pb collisions at $\sqrt{s_{NN}}$ = 2.76 TeV}
\author{Sushanta Tripathy}
\author{Arvind Khuntia}
\author{Swatantra~Kumar~Tiwari}
\author{Raghunath~Sahoo\footnote{Corresponding author: $Raghunath.Sahoo@cern.ch$}}
\affiliation{Discipline of Physics, School of Basic Sciences, Indian Institute of Technology Indore, Indore- 453552, INDIA}

\begin{abstract}
\noindent
In the continuation of our previous work, the transverse momentum ($p_T$) spectra and nuclear modification factor ($R_{AA}$) are derived using the relaxation time approximation of Boltzmann Transport Equation (BTE). The initial $p_T$-distribution used to describe $p+p$ collisions has been studied with the pQCD inspired power-law distribution, Hagedorn's empirical formula and with the Tsallis non-extensive statistical distribution. The non-extensive Tsallis distribution is observed to describe the complete range of the transverse momentum spectra. The Boltzmann-Gibbs Blast Wave (BGBW) distribution is used as the equilibrium distribution in the present formalism, to describe the $p_T$-distribution and nuclear modification factor in nucleus-nucleus collisions. The experimental data for Pb+Pb collisions at $\sqrt{s_{NN}}$ = 2.76 TeV at the Large Hadron Collider at CERN have been analyzed for pions, kaons, protons, $K^{*0}$ and $\phi$. It is observed that the present formalism while explaining the transverse momentum spectra up to 5 GeV/c, explains the nuclear modification factor very well up to 8 GeV/c in $p_T$ for all these particles except for protons.  $R_{AA}$ is found to be independent of the degree of non-extensivity, $q_{pp}$ after $p_T \sim$ 8 GeV/c.


\end{abstract}
\date{\today}
\maketitle 
\section{Introduction}
\label{intro}
The search for a deconfined state of quarks and gluons is a major goal of ongoing experiments based on relativistic heavy-ion collisions at high energies, like those at the Relativistic Heavy Ion Collider (RHIC) at the Brookhaven National Laboratory, USA and at the Large Hadron Collider (LHC) at CERN, Switzerland. These experiments are designed to create a plasma of quarks and gluons, called Quark-Gluon Plasma (QGP); which might have been formed after few micro-seconds of the Big Bang. The created matter is described by partonic degrees of freedom. The complete understanding of the properties of this newly created matter has been very challenging. Due to the parton (quarks and gluons) energy loss in the medium, suppression in particle yields is observed in nucleus-nucleus collisions relative to $p+p$ collisions, where the formation of a medium is usually not expected. Hence, the measurement of the suppression in particle yield is an ideal diagnostic means to probe the medium. The amount of suppression is generally measured with the help of nuclear modification factor, $R_{AA}$, which is defined as \cite{Adare:2008qa}:
\ba
\label{eq1}
 R_{AA}(p_{T})=\frac{(1/N_{AA}^{evt})d^{2}N_{AA}/dydp_{T}}{(\langle
N_{coll}\rangle/\sigma_{NN}^{inel})\times d^2\sigma_{pp}/dydp_T},
\ea
where $d^{2}N_{AA}/dydp_{T}$ is the yield and $N_{AA}^{evt}$ is the number of events in nucleus-nucleus (A+A) collisions. $\langle N_{coll}\rangle$ is the number of binary nucleon-nucleon collisions averaged over the impact parameter range of the corresponding centrality bin calculated by Glauber Monte-Carlo simulation \cite{Glauber:1970jm}. $\sigma_{NN}^{inel}$ is the inelastic cross section and $d^2\sigma_{pp}/dydp_T$ is the differential cross section for inelastic $p+p$ collisions. If A+A collisions are considered as mere superposition of scaled $p+p$ collisions, then $R_{AA}$ should always be unity. Deviation of $R_{AA}$ from unity indicates a medium modification. The observation of suppression in high-$p_T$ particle yields in Au+Au and Pb+Pb collisions at RHIC and LHC as compared to $p+p$ collisions \cite{Adler:2003qi,Aamodt:2010jd}, suggests the formation of a dense medium. Also, $R_{AA}$ can be represented as the ratio between the final distribution of particles ($f_{fin}$) and the initial particle distribution ($f_{in}$).

In this work, the initial distribution of the energetic particles is represented by the Tsallis power law distribution characterized by the Tsallis $q_{pp}$ parameter and the Tsallis temperature $T_{pp}$, remembering the fact that their genesis is due to very hard scatterings. Here the parameter $q_{pp}$, represents the degree of non-extensivity or in other words the degree of deviation of the system from a thermalized or equilibrated system, which is usually described by the well-known Boltzmann-Gibbs (BG) statistical mechanics. We plug in the initial distribution ($f_{in}$) in the Boltzmann Transport Equation (BTE) and solve it with the help of the Relaxation Time Approximation (RTA) of the collision term to find out the final distribution ($f_{fin}$). The final distribution includes both equilibrium and Tsallis distribution. In our earlier work \cite{Tripathy:2016hlg}, we have used the Boltzmann-Gibbs distribution as an equilibrium distribution function to study the $R_{AA}$ using the secondaries produced in A+A collisions. As BG distribution only describes the $p_T$-spectra in A+A collisions upto moderate $p_T$ and BGBW has been quite helpful in describing the $p_T$-spectra up to higher $p_T$, we use the latter distribution as the equilibrium distribution in the present formalism. The Boltzmann-Gibbs Blast Wave function has built-in radial collective flow. Now, the $R_{AA}$ is expressible in terms of $q_{pp}$, $T_{pp}$, $<\beta_{r}>$ and relaxation time $\tau$, which can be computed and compared with the experimental observations.

The paper is organized as follows. In section \ref{raa}, the nuclear modification factor is derived using the Relaxation Time Approximation of the Boltzmann Transport Equation. In section \ref{results}, fits to the experimental data ($p_T$ and $R_{AA}$ spectra) using the proposed model along with results and discussions are presented. Finally, we summarize our findings in section \ref{summary}.
 
\noindent
\section{Nuclear Modification Factor in Relaxation time approximation (RTA)}
\label{raa}
The evolution of the particle distribution owing to its interaction with the medium particles can be studied through the Boltzmann transport equation,
\begin{eqnarray}
\label{eq2}
 \frac{df(x,p,t)}{dt}=\frac{\partial f}{\partial t}+\vec{v}.\nabla_x
f+\vec{F}.\nabla_p
f=C[f],
\end{eqnarray}
where $f(x,p,t)$ is the distribution of particles which depends on position, momentum and time. {\bf v} is the velocity and {\bf F} is the external force. $\nabla_x$ and $\nabla_p$ are the partial derivatives with respect to position and momentum, respectively. $C[f]$ is the collision term which encodes the interaction of the probe particles with the medium. Earlier, BTE has also been used in the relaxation time approximation to study the time evolution of temperature fluctuation in a non-equilibrated system \cite{trambak-rns} and also for studying the $R_{AA}$ of various light and heavy flavours at RHIC and LHC energies \cite{Tripathy:2016hlg}. 

Assuming homogeneity of the system ($\nabla_x f=0$) and absence of external forces (F$=$0), the second and third terms of the Eq. \ref{eq2} become zero and reduces to,
\ba
 \label{eq3}
  \frac{df(x,p,t)}{dt}=\frac{\partial f}{\partial t}=C[f].
\ea

In the relaxation time approximation \cite{balescu,Florkowski:2016qig}, the collision term can be expressed as :
\ba
\label{eq4}
 C[f] =-\frac{f-f_{eq}}{\tau},
 \label{colltermrta}
\ea
where $f_{eq}$ is Boltzmann local equilibrium distribution characterized by a temperature $T$. $\tau$ is the relaxation time, the time taken by a non-equilibrium system to reach equilibrium. Using Eq. \ref{colltermrta}, Eq. \ref{eq3} becomes 
\ba
 \label{eq5}
  \frac{\partial f}{\partial t}=-\frac{f-f_{eq}}{\tau}.
\ea
Solving the above equation in view of the initial conditions {\it i.e.} at $t=0, f=f_{in}$ and at $t=t_f, f=f_{fin}$, leads to,
\ba
 \label{eq6}
 f_{fin}=f_{eq}+(f_{in}-f_{eq})e^{-\frac{t_f}{\tau}},
\ea
where $t_f$ is the freeze-out time. Using Eq. \ref{eq6}, the nuclear modification factor can be expressed as,
\ba
\label{eq7}
R_{AA}=\frac{f_{fin}}{f_{in}}=\frac{f_{eq}}{f_{in}}+\left ( 1-\frac{f_{eq}}{f_{in}}\right )e^\frac{-t_{f}}{\tau}.
\ea
Eq. \ref{eq7} is the derived nuclear modification factor after incorporating relaxation time approximation, which is the basis of our analysis in the present paper. It involves the Tsallis non-extensive distribution function as the initial distribution and the BGBW function as the equilibrium distribution. 

Here, we take the Boltzmann-Gibbs Blast Wave (BGBW) function as $f_{eq}$, which is given by 
\ba
\label{boltz_blast}
f_{eq} = D \int_0^{R_{0}} r\;dr\;K_1\Big(\frac{m_T\;cosh\rho}{T}\Big)I_0\Big(\frac{p_T\;sinh\rho}{T}\Big),
\ea
where $D = \displaystyle \frac{gVm_T}{2\pi^2}$. Here $g$ is the degeneracy factor, $V$ is the system volume, and $m_{\rm T}=\sqrt{p_T^2+m^2}$ is the transverse mass, $K_{1}\displaystyle\Big(\frac{m_T\;{\cosh}\rho}{T}\Big)$ and $I_0\displaystyle\Big(\frac{p_T\;{\sinh}\rho}{T}\Big)$ are the modified Bessel's functions and are given by
\begin{widetext}
\ba
\centering
K_1\Big(\frac{m_T\;{\cosh}\rho}{T}\Big)=\int_0^{\infty} {\cosh}y\;{\exp}\Big(-\frac{m_T\;{\cosh}y\;{\cosh}\rho}{T}\Big)dy,
\ea
\ba
\centering
I_0\Big(\frac{p_T\;{\sinh}\rho}{T}\Big)=\frac{1}{2\pi}\int_0^{2\pi} exp\Big(\frac{p_T\;{\sinh}\rho\;{\cos}\phi}{T}\Big)d\phi,
\ea
\end{widetext}
where $\rho$ is a parameter given by $\rho=tanh^{-1}\beta_r$, with $\beta_r=\displaystyle\beta_s\;\Big(\xi\Big)^n$ \cite{Schnedermann:1993,Braun:1996}. $\beta_s$ is the maximum surface velocity and $\xi=\displaystyle\Big(r/R_0\Big)$, with $r$ as the radial distance. This is similar to the Hubble expansion of the universe ($v = Hr$, H is the Hubble constant). In the blast-wave model
the particles closer to the center of the fireball move slower than the ones on the edges. The average of the transverse velocity can be evaluated as \cite{Adcox:2004} 
\ba
<\beta_r> =\frac{\int \beta_s\xi^n\xi\;d\xi}{\int \xi\;d\xi}=\Big(\frac{2}{2+n}\Big)\beta_s.
\ea
In our calculation we use a linear velocity profile, ($n=1$) and $R_0$ is the maximum radius of the expanding source at freeze-out ($0<\xi<1$). In this analysis, the initial distribution is parameterized using three different distributions: (i) the pQCD motivated power-law distribution which is given as,
\ba
\label{eq11}
f_{in}=\frac{gV}{(2\pi)^2} m_T
\left[{\frac{m_T}{T_{pp}}}\right]^{-n},
\ea
(ii) the distribution proposed by Hagedorn, which is a combination of exponential distribution for low-$p_T$ and power-law distribution for high-$p_T$, is expressed as,
\ba
\label{eq12}
f_{in}=\frac{gV}{(2\pi)^2} m_T
\left[1+{\frac{m_T}{T_{pp}}}\right]^{-n},
\ea
and (iii) the thermodynamically consistent non-extensive Tsallis distribution \cite{worku}
\ba
\label{eq13}
f_{in}=\frac{gV}{(2\pi)^2} m_T
\left[1+{(q_{pp}-1)}{\frac{m_T}{T_{pp}}}\right]^{-\frac{q_{pp}}{q_{pp}-1}}.
\ea
Using all the above distributions, we have analyzed the $R_{AA}$ spectra and it is observed that the pQCD inspired power-law distribution could explain the high transverse momentum part but fails in low momentum range. Thus, we have used the Tsallis distribution to obtain the expression for the final distribution and nuclear modification factor. The thermodynamically consistent Tsallis distribution is used for studying the particle distributions stemming from the proton-proton collisions as discussed in Ref. \cite{worku}. $T_{pp}$ is the Tsallis temperature and $q_{pp}$ is the non-extensive parameter, which measures the degree of deviation from equilibrium.

Using Eqs. \ref{boltz_blast} and \ref{eq13}, the final distribution can be expressed as, 

%\begin{eqnarray}
%\label{eq12}
%\nonumber f_{fin}=\int_0^{R_{0}} r\;dr\;K_1\Big(\frac{m_T\;cosh\rho}{T}\Big)I_0\Big(\frac{p_T\;sinh\rho}{T}\Big)} +\\\left(\frac{\left[1+{(q-1)}{\frac{m_T}{T}}\right]^{-\frac{q}{q-1}}}{\left[1+{(q_{pp}-1)}{\frac{m_T}{T}}\right]^{-\frac{q_{pp}}{q_{pp}-1}}}- \frac{\int_0^{R_{0}} r\;dr\;K_1\Big(\frac{m_T\;cosh\rho}{T}\Big)I_0\Big(\frac{p_T\;sinh\rho}{T}\Big)}{\left[1+{(q_{pp}-1)}{\frac{m_T}{T}}\right]^{-\frac{q_{pp}}{q_{pp}-1}}}\right )e^\frac{-t_{f}}{\tau}
%\end{eqnarray}

\begin{widetext}
\ba
\label{eq14}
\nonumber f_{fin} = D \Big\{\int_0^{R_{0}} r\;dr\;K_1\Big(\frac{m_T\;{\cosh}\rho}{T}\Big)I_0\Big(\frac{p_T\;{\sinh}\rho}{T}\Big)+\\\left(\frac{1}{2}\left[1+{(q_{pp}-1)}{\frac{m_T}{T_{pp}}}\right]^{-\frac{q_{pp}}{q_{pp}-1}}-\int_0^{R_{0}}r\;dr\;K_1\Big(\frac{m_T\;{\cosh}\rho}{T}\Big)I_0\Big(\frac{p_T\;{\sinh}\rho}{T}\Big)\right)e^\frac{-t_{f}}{\tau}\Big\}.%\right)e^\frac{-t_{f}}{\tau}
\ea
\end{widetext}
Using Eqs. \ref{boltz_blast} and \ref{eq14} (both for mid-rapidity and for zero chemical potential) in Eq. \ref{eq7},
nuclear modification factor can be expressed as,
\begin{widetext}
\ba
\label{eq15}
R_{AA}=\frac{f_{fin}}{f_{in}}=\frac{\int_0^{R_{0}} r\;dr\;K_1\Big(\frac{m_T\;cosh\rho}{T}\Big)I_0\Big(\frac{p_T\;sinh\rho}{T}\Big)}{\frac{1}{2}\left[1+{(q_{pp}-1)}{\frac{m_T}{T_{pp}}}\right]^{-\frac{q_{pp}}{q_{pp}-1}}}+\left(1-\frac{\int_0^{R_{0}} r\;dr\;K_1\Big(\frac{m_T\;cosh\rho}{T}\Big)I_0\Big(\frac{p_T\;sinh\rho}{T}\Big)}{\frac{1}{2}\left[1+{(q_{pp}-1)}{\frac{m_T}{T_{pp}}}\right]^{-\frac{q_{pp}}{q_{pp}-1}}}\right )e^\frac{-t_{f}}{\tau}.
\ea
\end{widetext}
Here, $T_{pp}$ and $q_{pp}$ are extracted from the best fit to the particle spectra in $p+p$ collisions. The Eqs. \ref{eq14} and \ref{eq15} are used to fit the experimental results as discussed in the following section.

\begin{figure}
\includegraphics[height=15em]{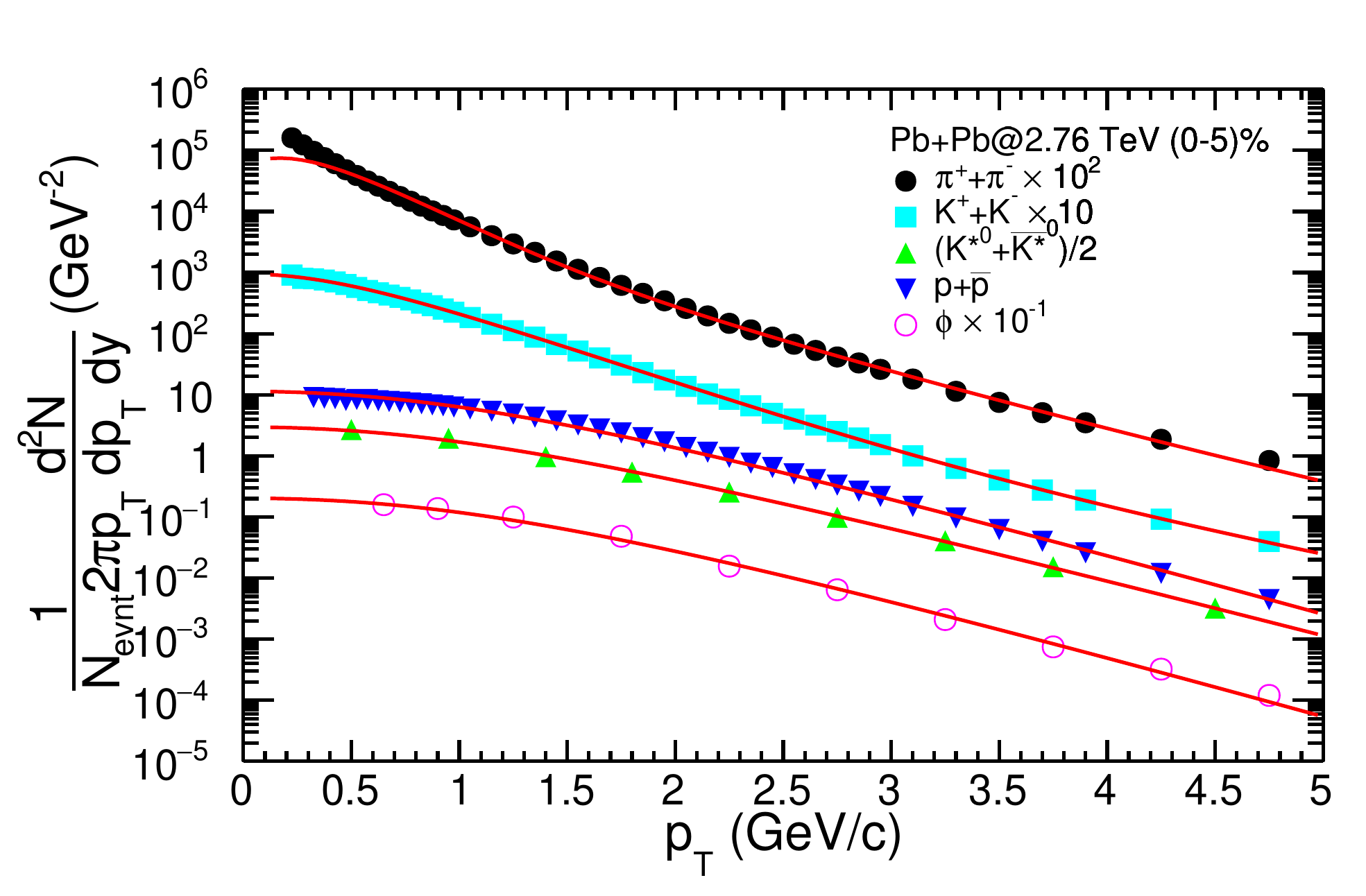}%use pdflatex for pdf
\caption[]{(Color online) The invariant yield of various particles in most central Pb+Pb collisions at $\sqrt{s_{NN}} =$ 2.76 TeV at mid-rapidity. The fitted lines are the final distributions given by Eq. \ref{eq14}.}
\label{fig1}
\end{figure}

\begin{figure}
\includegraphics[height=15em]{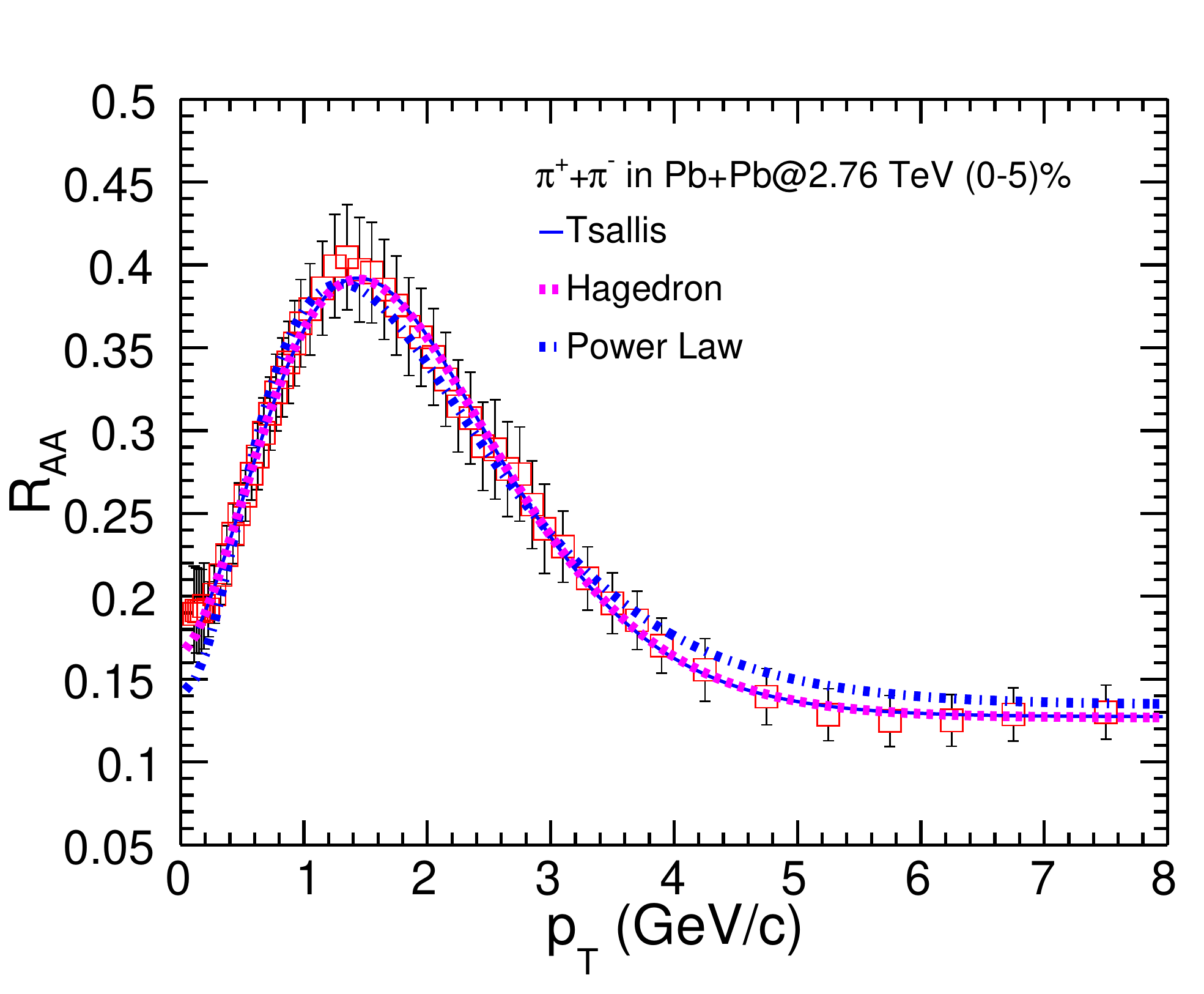}
\caption[]{(Color online) Fitting of $R_{AA}$ spectra for pions in most central Pb+Pb collisions at $\sqrt{s_{NN}} = 2.76$ TeV, using three different initial distributions. The blue dotted line shows the fitting using the power-law distribution (Eq. \ref{eq11}), the magenta dotted line shows the fitting using the Hagedorn distribution (Eq. \ref{eq12}) and the blue line shows the fitting using the thermodynamically consistent Tsallis distribution (Eq. \ref{eq13}).}
\label{fig2}
\end{figure}

\section{Results and Discussions}
\label{results}
We now proceed to the more detailed analysis of the experimental data with the proposed formulation. Keeping all the parameters free, we fit the spectra for different particles in most central Pb+Pb collisions using TMinuit class available in ROOT library \cite{root} to get a convergent solution. The convergent solution is obtained by $\chi ^2$-minimization technique. Fig. \ref{fig1} shows the invariant $p_T$-spectra of $\pi^++\pi^-$, $K^++K^-, (K^{*0}+\bar{K^{*0}})/2, p+\bar{p}$ and $\phi$ for the most central Pb+Pb collisions at $\sqrt{s_{NN}} =$ 2.76 TeV at mid-rapidity. The fitted lines are the expectations from the final distribution given by Eq. \ref{eq14}. The Eq. \ref{eq14} describes the experimental data very well up to $p_T$ = 5 GeV/c for all the particles with a very good $\chi^2$/ndf except for $p+\bar{p}$ and $\pi^++\pi^-$, for which we do not get a good $\chi^2$/ndf after $p_T$ = 3 GeV/c. Here $T_{pp}$, $q_{pp}$, $<\beta_r>$ and ${t_f}/{\tau}$ are the fitting parameters for the experimental data of transverse momentum ($p_{T}$) spectra. The equilibrium temperature, $T$ is fixed to 160 MeV throughout the analysis. 

\begin{figure}
\includegraphics[height=15em]{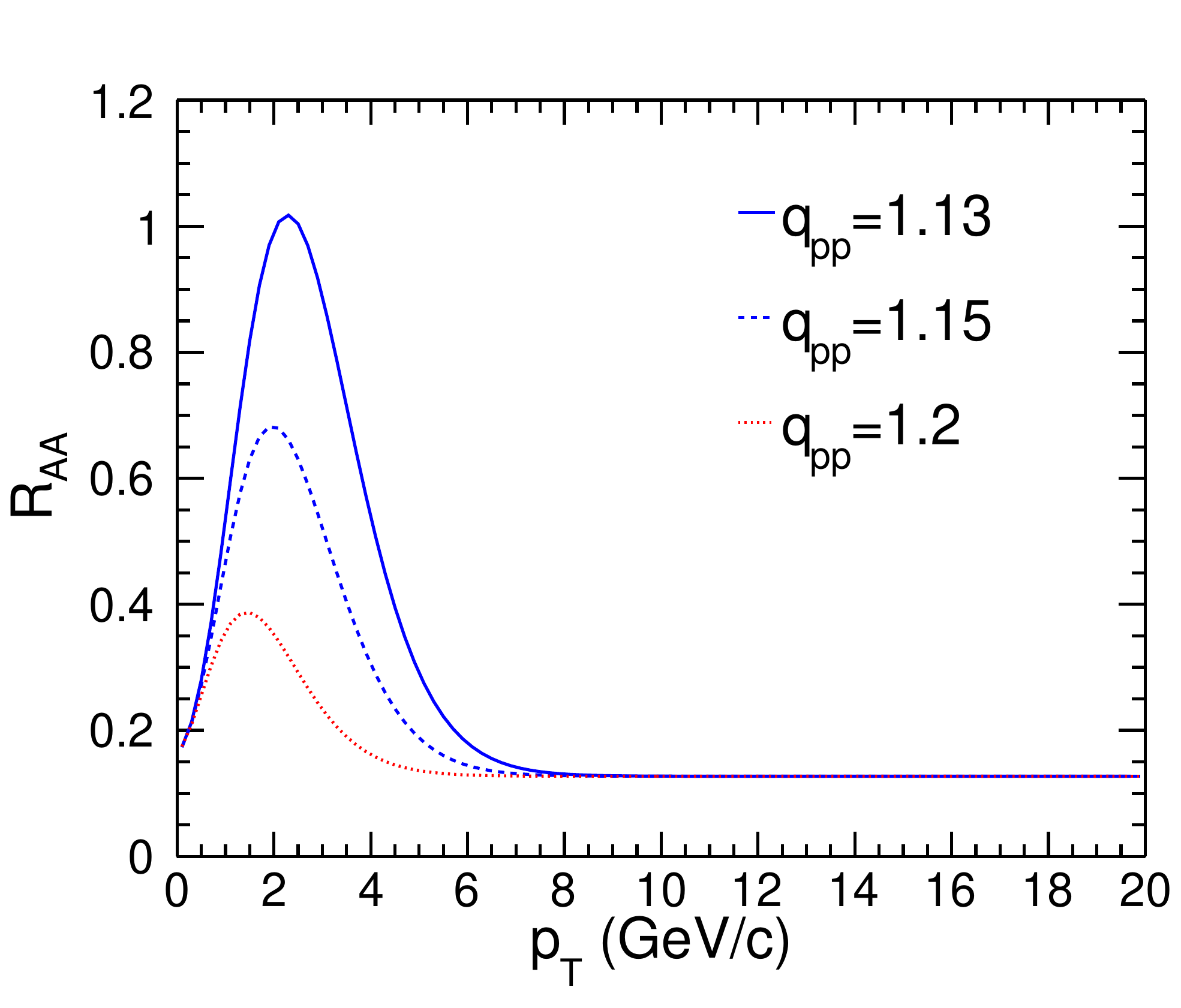}
\caption[]{(Color online) $R_{AA}$ spectra for pions as a function of the non-extensive parameter, $q_{pp}$ using Eq. \ref{eq15}. Here m = 0.139 GeV/$c^{2}$, $T$ = 0.16 GeV, $T_{pp}$ = 0.108 GeV, $t_{f}/\tau$ = 2.06 and $\beta_{r}$ = 0.501 are taken.}
\label{R_AA-q}
\end{figure}

\begin{figure}
\includegraphics[height=15em]{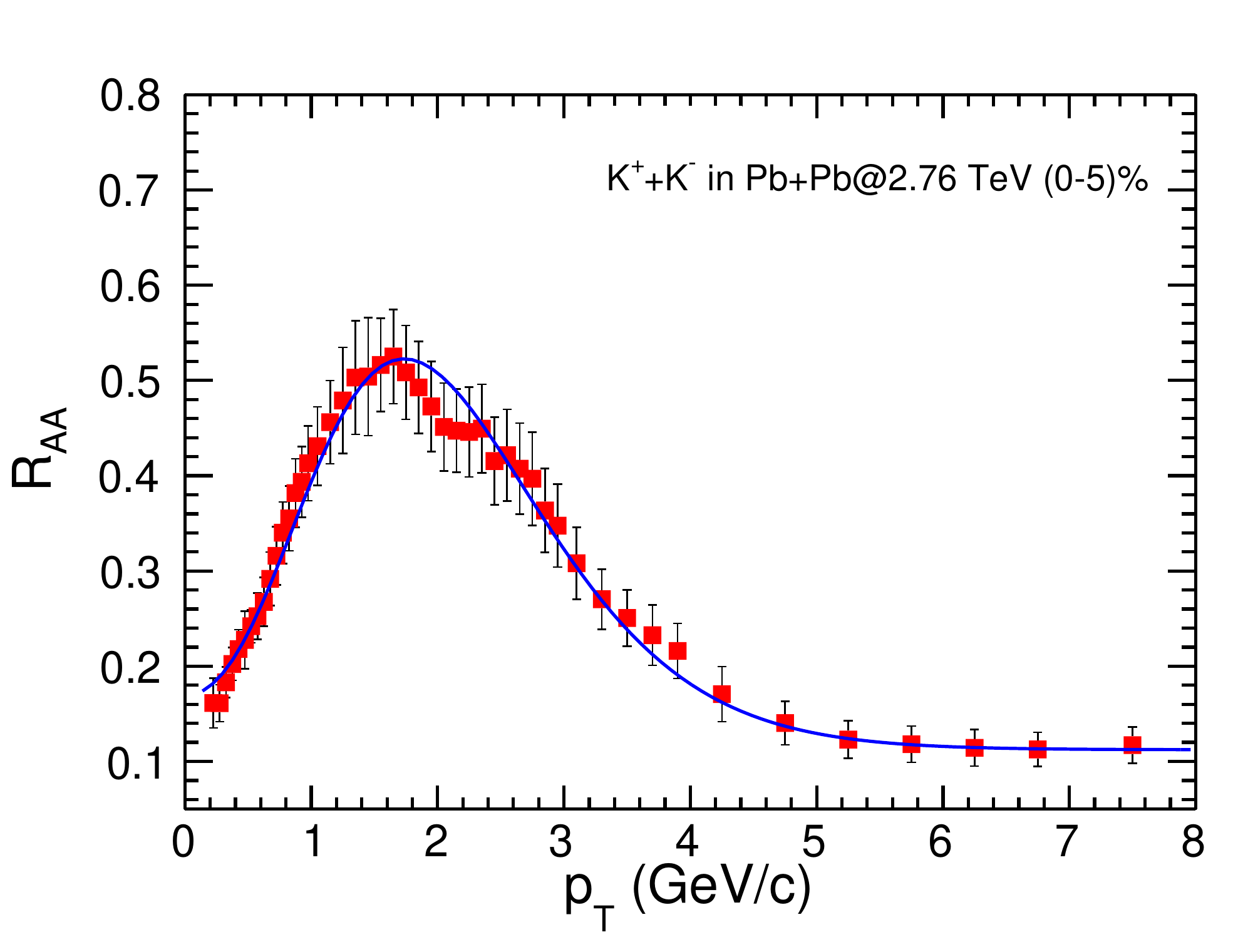}
\caption[]{(Color online) $R_{AA}$ spectra for kaons \cite{Abelev:2014laa} in most central Pb+Pb collisions at $\sqrt{s_{NN}}$ = 2.76 TeV. The solid line shows the agreement of the present formalism in describing the experimental data.}
\label{fig3}
\end{figure}

\begin{figure}
\includegraphics[height=15em]{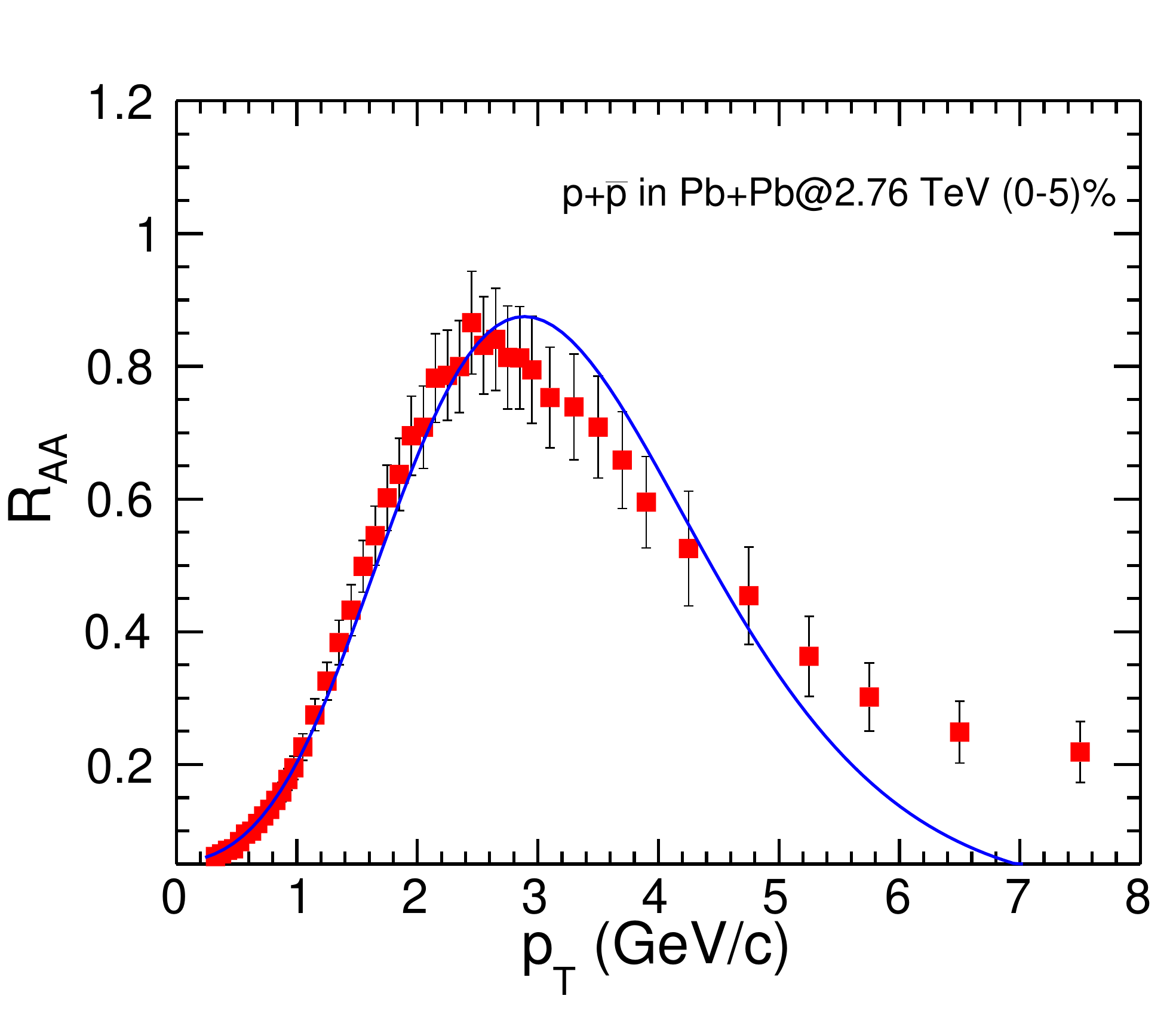}
\caption[]{(Color online) $R_{AA}$ spectra for protons \cite{Abelev:2014laa} in most central Pb+Pb collisions at $\sqrt{s_{NN}} = 2.76$ TeV. The solid line shows the agreement of the present formalism in describing the experimental data.}
\label{fig4}
\end{figure} 

\begin{figure}
\includegraphics[height=15em]{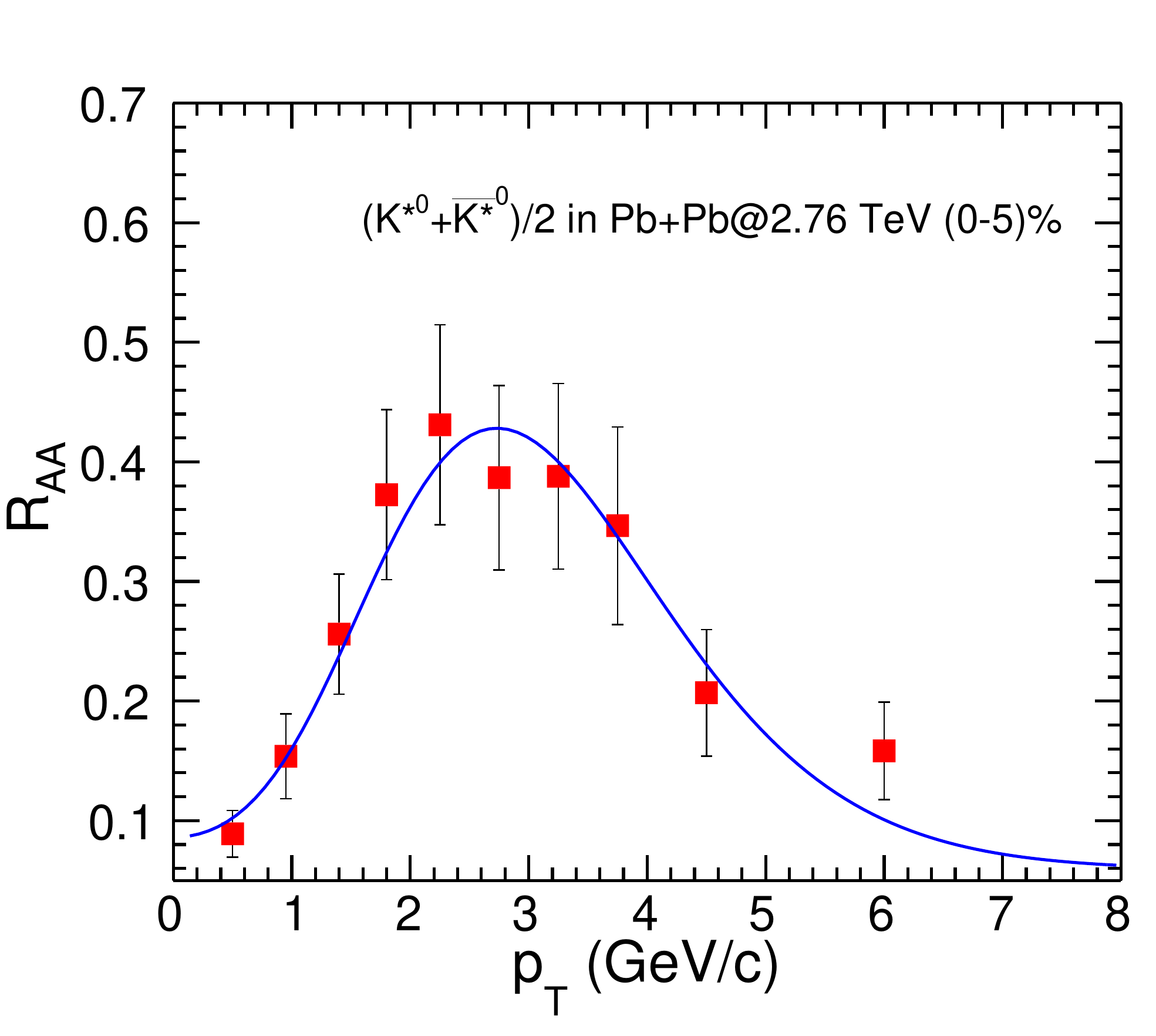}
\caption[]{(Color online) $R_{AA}$ spectra for $(K^{*0}+\bar{K^{*0}})/2$ \cite{Adam:2017zbf} in most central Pb+Pb collisions at $\sqrt{s_{NN}}$ = 2.76 TeV. The solid line shows the agreement of the present formalism in describing the experimental data.}
\label{fig5}
\end{figure}
Fig. \ref{fig2} shows $R_{AA}$ as a function of $p_T$ for pions in most central Pb-Pb collisions at $\sqrt{s_{NN}}$ = 2.76 TeV. Here, we fit the experimental data by taking the various types of initial distributions as mentioned above in order to check the suitability of the initial distributions used as $f_{in}$ in the definition of $R_{AA}$. It is observed that the pQCD inspired power-law distribution could explain the high transverse momentum part but fails in low momentum range compared to other two distributions. This is expected as the high-$p_T$ contribution mostly comes from hard scatterings, which are described by pQCD. The distribution proposed by Hagedorn behaves as an exponential distribution in low-$p_T$ and the power-law distribution in high-$p_T$ domains, explains $R_{AA}$ for the complete range of $p_T$. The thermodynamically consistent Tsallis distribution is also used to fit $R_{AA}$ in comparison with the above mentioned initial distributions. We find that using the thermodynamically consistent Tsallis distribution as an initial distribution, the $R_{AA}$ spectra are explained successfully. So, considering the whole $p_T$-range and taking the non-extensive statistics as an initial distribution, we further proceed towards studying $R_{AA}$ of light flavours and resonances in Pb+Pb collisions at $\sqrt{s_{NN}} = 2.76$ TeV at the LHC. We also notice that using the BGBW function as $f_{eq}$ in the definition of $R_{AA}$ gives a very good description in the whole $p_T$-range, while our earlier version of $R_{AA}$ formulation, which uses Boltzmann-Gibbs distribution as the equilibrium distribution, fails at the low-$p_T$ domain. This suggests that the collective flow plays an important role in the study of  $R_{AA}$ spectra. Further, in the description of $R_{AA}$ spectra of all other particles, we use the non-extensive Tsallis distribution function as $f_{in}$. This is because, in addition to a better description of $p_T$-spectra in $p+p$ collisions and the $R_{AA}$-spectra in nucleus-nucleus collisions in the present formalism, it also gives other thermodynamical properties of the system, which are quite useful in characterizing the matter formed at this energy.

\begin{figure}
\includegraphics[height=15em]{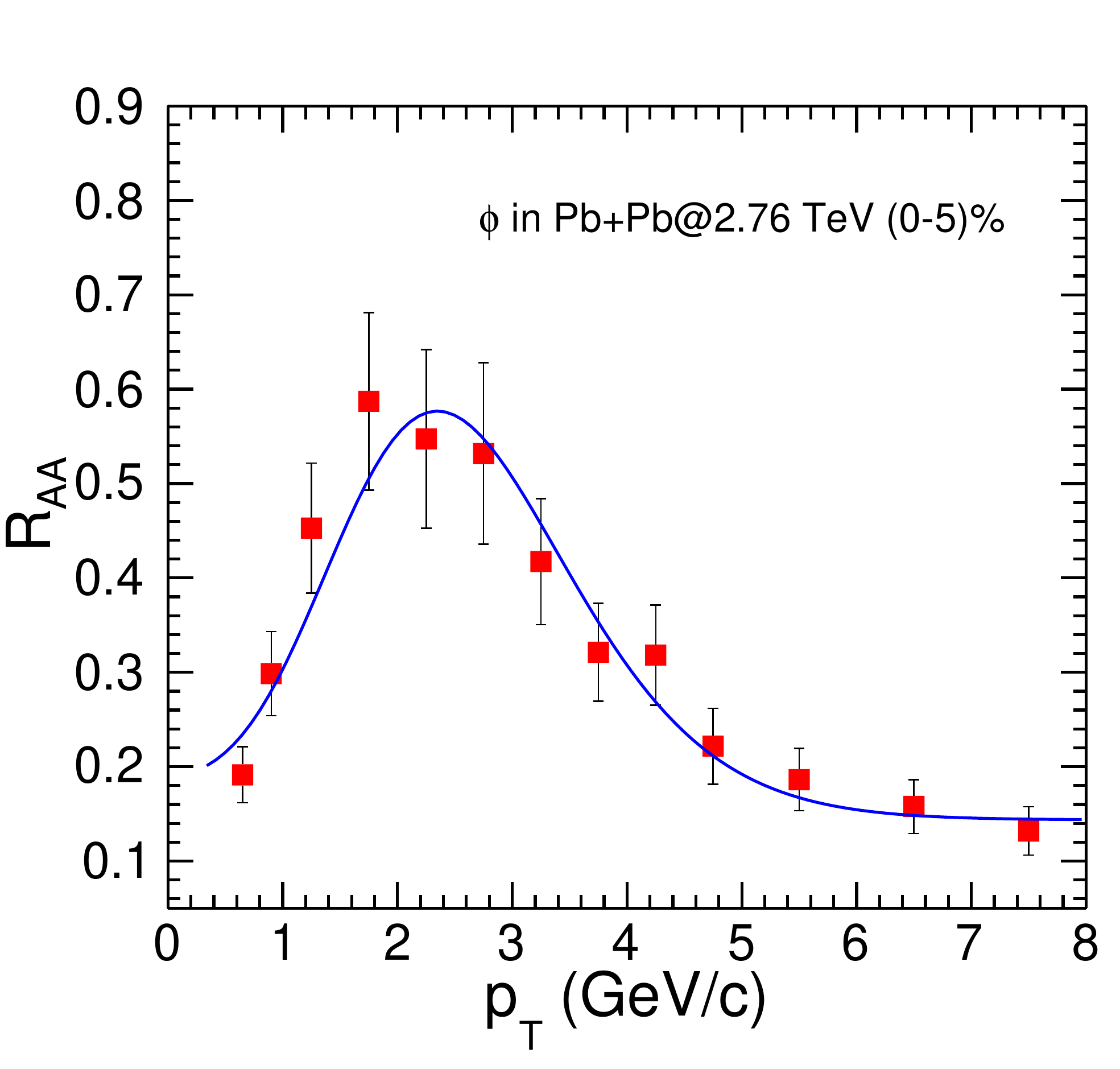}
\caption[]{(Color online) $R_{AA}$ spectra for $\phi$ \cite{Adam:2017zbf} in most central Pb+Pb collisions at $\sqrt{s_{NN}} = 2.76$ TeV. The solid line shows the agreement of the present formalism in describing the experimental data.}
\label{fig6}
\end{figure}
Figure \ref{R_AA-q} shows the variation of nuclear modification factor as a function of $p_{T}$ for different values of non-extensive parameter $q_{pp}$ following Eq. \ref{eq15}. Here m = 0.139 GeV/$c^{2}$, $T$ = 0.16 GeV, $T_{pp}$ = 0.108 GeV, $t_{f}/\tau$ = 2.06 and $<\beta_{r}>$ = 0.501. It is observed that, the $R_{AA}$ value decreases with increase in $q_{pp}$, which suggests that when the initial distribution remains closer to equilibrium (lower the value of $q_{pp}$), the suppression becomes less. This observation goes inline with our previous observation \cite{Tripathy:2016hlg}. Also, it is observed from Fig. \ref{R_AA-q} that the non-extensive parameter dependence on $R_{AA}$ spectra is only seen up to $p_T \sim$ 8 GeV/c. The flatness in $R_{AA}$, which is seen in higher-$p_T$, is observed to shift towards lower-$p_T$ for higher $q_{pp}$-values. These are very important observations. 
%which tells that the high-$p_T$ particles get less time to interact within the medium and hence have lesser tendency of equilibration. This observation becomes independent of the non-extensivity of the system at high-$p_T$, as is observed from the above figure. }

\begin{figure}
\includegraphics[height=15em]{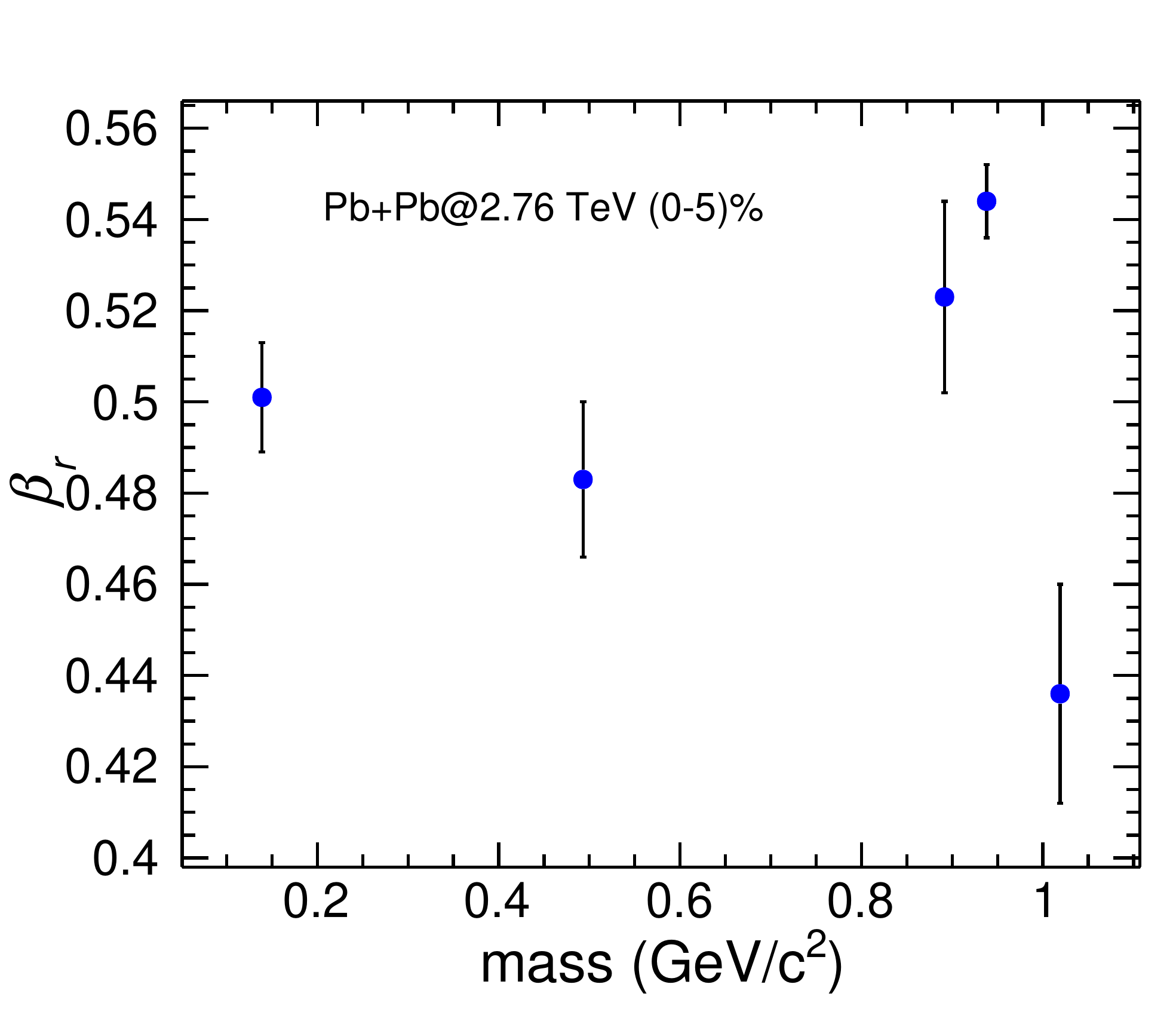}
\caption[]{(Color online) Radial flow ($<\beta_r>$) as a function of particle mass.}
\label{fig7}
\end{figure}
Figures \ref{fig3}, \ref{fig4}, \ref{fig5}, and \ref{fig6} show the $R_{AA}$-spectra of $K^++K^-$, $p+\bar{p}$, ($K^{*0}+\bar{K^{*0}}$)/2 and $\phi$ for the most central Pb+Pb collisions at $\sqrt{s_{NN}}$ = 2.76 TeV, respectively. The spectra are fitted to the the function for $R_{AA}$ given by Eq. \ref{eq15}, which is obtained in the present formalism of BTE with RTA and BGBW function as the equilibrium distribution. The extracted parameters are enlisted in Table \ref{table}. As could be observed from the above figures the present formalism explains the nuclear modification factor very well up to 8 GeV/c for all the particles 
except $p+\bar{p}$, for which it is explained up to 5 GeV/c. This is because $p+\bar{p}$ shows an enhancement in the yield after $p_T \sim$ 3 GeV/c, which is not seen for other particles \cite{Adam:2017zbf}.

\begin{figure}
\includegraphics[height=15em]{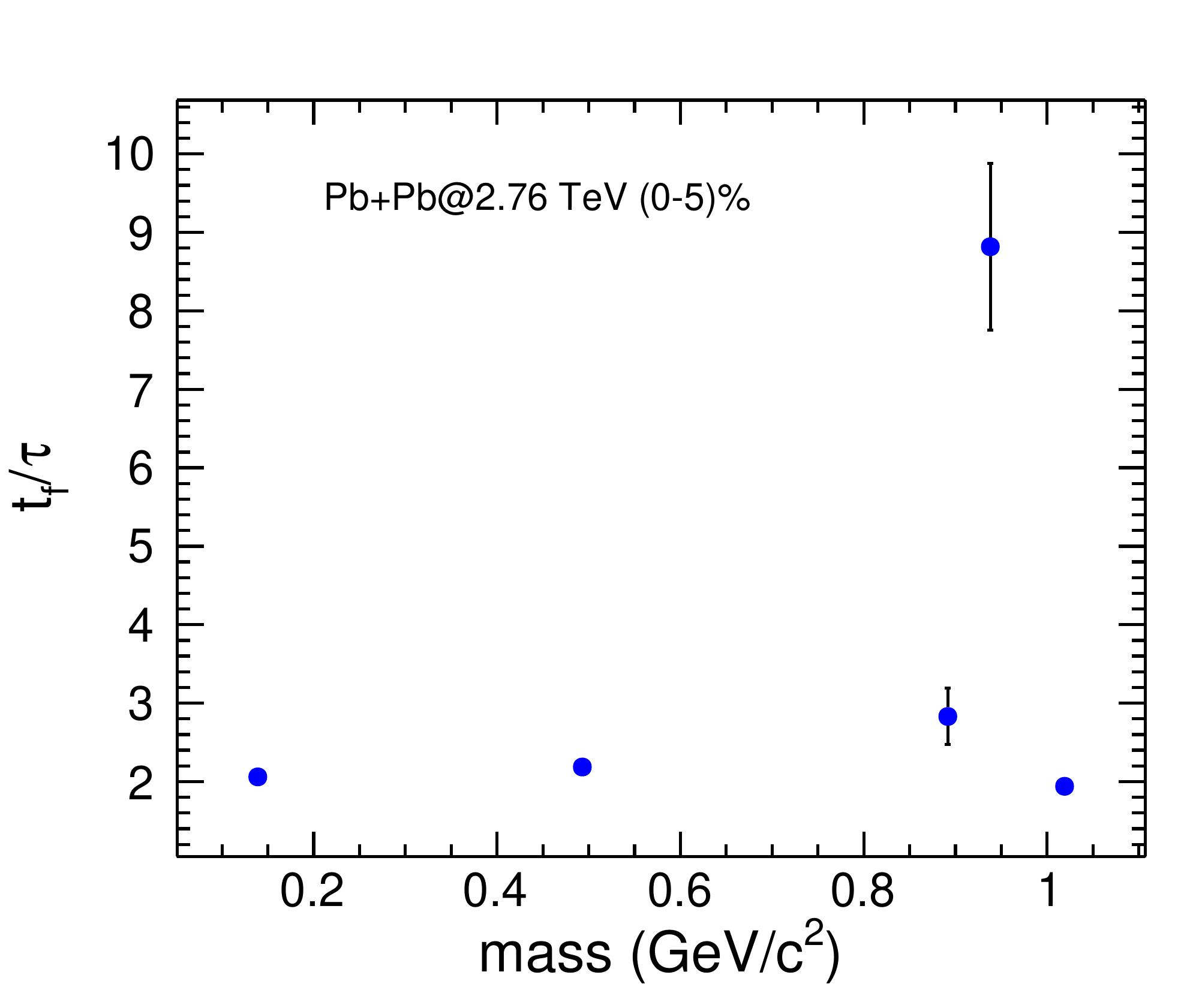}
\caption[]{(Color online) The ratio of freeze-out time to relaxation time ($t_f/\tau$) as a function of particle mass.}
\label{fig8}
\end{figure}

In Fig. \ref{fig7}, we show the variation of radial flow with the mass of the particles, extracted from the $R_{AA}$ spectra in the present formulation. Higher mass particles seem to have lower flow velocity, which goes in line with the hydrodynamic behavior of collectivity in these systems. However, as can be seen here, the ($K^{*0}+\bar{K^{*0}}$)/2 and $p+\bar{p}$ do not follow the same trend.

Figure \ref{fig8} shows the variation of the ratio of freeze-out time ($t_f$) to the relaxation time ($\tau$) with the mass of the particles. This ratio has been extracted from the fitting to the $R_{AA}$ spectra. $t_f/\tau$ is almost independent of the particle mass, except for the protons and anti-protons. Although this does not go in line with the intuitive expectations, as the degrees of freedom are shared between various parameters like, $<\beta_r>$, $q_{pp}, T_{pp}$ and $t_f/\tau$, one needs to understand the interplay of these parameters.     

\section{Summary and Conclusion}
\label{summary}
In this work, we have made an attempt to explain the transverse momentum spectra and nuclear modification factor of various particles produced in central Pb+Pb collisions at $\sqrt{\mathrm{s}_{NN}}$= 2.76 TeV at the LHC, in a single approach. This formalism uses the Boltzmann Transport Equation in Relaxation Time Approximation, where we have taken a thermodynamically consistent Tsallis non-extensive distribution function as the initial distribution of the particle momenta. Using the BTE, we study the time evolution of the initial distribution function to find the final distribution function of the particles. In this approach, we have used the Boltzmann-Gibbs Blast Wave function as the equilibrium distribution in the nucleus-nucleus collisions, where collective radial flow plays an important role in describing the transverse momentum distribution.  In summary,
\begin{enumerate}
\item In this formalism, we find that the final distribution function describes the transverse momentum spectra and the nuclear modification factor of pions, kaons, protons, $K^{*0}$ and $\phi$ up to considerably high $p_T$.
\item The extracted radial flow seems to be mass dependent and favours a hydrodynamic behavior except for ($K^{*0}+\bar{K^{*0}}$)/2 and $p+\bar{p}$, which needs further studies. 
\item $R_{AA}$ is found to be independent of the degree of non-extensivity, $q_{pp}$ after $p_T \sim$ 8 GeV/c. The flatness in $R_{AA}$, which is seen in higher-$p_T$, is observed to shift towards lower-$p_T$ for higher $q_{pp}$-values.
\item The non-extensivity parameter, $q_{pp}$ is mass dependent and it decreases for higher mass particles. Higher mass particles have a tendency of fast equilibration.
\item The inclusion of radial flow, $<\beta_r>$ in the theory, favours the non-extensivity, as is expected intuitively. This is seen when we compare the present results with our earlier findings \cite{Tripathy:2016hlg}.

\item The ratio, $t_f/\tau$ seems to be independent of particle mass, except for protons and anti-protons. Although this does not go in line with the intuitive expectations, where a decrease of $t_f/\tau$ with increase in the mass is expected \cite{Tripathy:2016hlg}, as the degrees of freedom are shared with other parameters, a microscopic understanding is thus required for a clear picture of the interplay of radial flow, relaxation time and the non-extensivity of the system.

\end{enumerate}

\section*{Acknowledgements}
ST acknowledges the financial support by DST INSPIRE program of Govt. of India. The authors would like to put on record the helps from Dr. Santosh K. Das and Dr. Trambak Bhattacharyya for having read the article and giving valuable comments.

\begin{table*}
\centering
%\begin{center}
\caption {${\chi ^2}/{ndf}$ and different extracted parameters
after fitting Eq.~\ref{eq15} to the $R_{AA}$ data of different
particles for most central Pb+Pb collisions at $\sqrt{\mathrm{s}_{NN}}$= 2.76 TeV.}

\noindent\begin{tabular}{ |c|c|c|c|c|c| }
\hline
\multicolumn{6}{|c|}{{\bf Pb+Pb 2.76 TeV}}\\
\hline
Particle & ${\chi ^2}/{ndf}$ & $<\beta_r>$ & $t_f/\tau$ & $q_{pp}$  & $T_{pp}$ (GeV) \\
\hline
$\pi ^+ + \pi ^-$         & 0.064 & 0.501 $\pm$ 0.012 	& 2.060 $\pm$ 0.054  & 1.200 $\pm$ 0.140 & 0.108 $\pm$ 0.009  \\
$K^+ + K^-$              & 0.290 &	0.483 $\pm$ 0.017	& 2.186 $\pm$ 0.091  & 1.200 $\pm$ 0.012 & 0.053 $\pm$ 0.015  \\
($K^{*0}+\bar{K^{*0}}$)/2  & 0.744  & 0.523 $\pm$ 0.021	& 2.830 $\pm$ 0.359  & 1.161  $\pm$ 0.019 & 0.051 $\pm$ 0.014  \\
$p+\bar p$    	        & 1.531 & 0.544 $\pm$ 0.008	& 8.817 $\pm$ 1.065  & 1.182 $\pm$ 0.006 & 0.028 $\pm$ 0.001  \\
$\phi$                    	& 0.869 &	0.436 $\pm$ 0.024	& 1.940 $\pm$ 0.091  & 1.140 $\pm$ 0.012 & 0.031 $\pm$ 0.005  \\
\hline
\end{tabular}
\label{table}
%\end{center}

%\end{table}

%\begin{table}[H]
%\begin{center}
\end{table*}

\end{document}